\documentclass[aps,nobibnotes,prb,superscriptaddress,amsmath,amssymb,amsfonts,twocolumn]{revtex4-1}
\usepackage{array,multirow,graphicx}
\usepackage{color}
\usepackage{ulem}
\usepackage[latin1]{inputenc}
\usepackage{caption}
\usepackage{scrextend}
\usepackage{hyperref}
\usepackage{float}
\usepackage{subcaption}

\def\sss{\scriptscriptstyle\rm}
\def\br{{\bf r}}
\def\bq{{\bf q}}
\def\bG{{\bf G}}
\def\xc{_{\sss XC}}
\def\bn{\boldsymbol{\nabla}}
\def\ext{_{\rm ext}}

\begin{document}

\normalem 

\title{The generalized gradient approximation kernel in time-dependent density functional theory}
\author{N. Singh}
\affiliation{Max-Planck-Institut f\"ur Mikrostrukturphysik, Weinberg 2, D-06120 Halle, Germany.}
\affiliation{Department of Physics, Indian Institute of Technology-Roorkee, 247667 Uttarakhand, India.}
\author{P. Elliott}
\affiliation{Max-Planck-Institut f\"ur Mikrostrukturphysik, Weinberg 2, D-06120 Halle, Germany.}
\author{T. Nautiyal}
\affiliation{Department of Physics, Indian Institute of Technology-Roorkee, 247667 Uttarakhand, India.}
\author{J.K. Dewhurst}
\affiliation{Max-Planck-Institut f\"ur Mikrostrukturphysik, Weinberg 2, D-06120 Halle, Germany.}
\author{S. Sharma}
\affiliation{Max-Planck-Institut f\"ur Mikrostrukturphysik, Weinberg 2, D-06120 Halle, Germany.}
\affiliation{Department of Physics, Indian Institute of Technology-Roorkee, 247667 Uttarakhand, India.}

\date{\today}

\begin{abstract}
A complete understanding of a material requires both knowledge of the excited states as well as of the ground state. In particular, the low energy excitations are of utmost importance while studying the electronic, magnetic, dynamical, and thermodynamical properties of the material. Time-Dependent Density Functional Theory (TDDFT), within the linear regime, is a successful \textit{ab-initio} method to access the electronic charge and spin excitations. However, it requires an approximation to the exchange-correlation (XC) kernel which encapsulates the effect of electron-electron interactions in the many-body system. In this work we derive and implement the spin-polarized XC kernel for semi-local approximations such as the adiabatic Generalized Gradient Approximation (AGGA). This kernel has a quadratic dependence on the wavevector, $\bq$, of the perturbation, however the impact of this on the electron energy loss spectra (EELS) is small. Although the GGA functional is good in predicting structural properties, it generality overestimates the exchange spin-splitting. This leads to higher magnon energies, as  compared to both ALDA and experiment. In addition, interaction with the Stoner spin-flip continuum is enhanced by AGGA, which strongly suppresses the intensity of spin-waves.

\end{abstract}

\maketitle

\section{INTRODUCTION}

Recent developments in the field of laser-induced spin-dynamics have opened up the world of \emph{femtomagnetism} \cite{U09}, whereby the spin degree-of-freedom is controlled using ultrafast laser pulses \cite{BMDB96}. As the name suggests, femtomagnetism concerns charge and spin dynamics on the femtosecond ($=10^{-15}$ s) time scale, corresponding to energies in the meV range. Electronic excitations in this energy regime can be clasified as either localized \emph{single-particle} like transitions, e.g. Stoner spin-flips, or those exhibiting a \emph{collective} nature, such as excitons or magnons \cite{Kittel}, which are spread over many atomic sites. These collective excitations with large wavelength occur at relatively lower energies as compared to the single-particle excitations and hence are important in studying the material's properties. To exploit the vast potential femtomagnetism offers, it is vital that we are able to accurately describe these collective excitations, in order to understand, and ultimately control, them. 
 
Theoretical studies of charge and spin excitations can be performed using either simple models like the Heisenberg model \cite{Heisenberg1928} or Landau-Lifshitz-Gilbert equation \cite{MBO18}, or computationally more demanding, parameter-free, \emph{ab-initio} methods. In contrast to \emph{ab-initio} methods, model based approaches are limited by their lack of generality, as they are usually tailored to study only specific problems and cannot be used universally for all materials.

Time-dependent density functional theory (TDDFT) \cite{RG84,SDG14,PE09} is an \emph{ab-initio} method which can predict the excited state properties of a material. Since its theoretical foundation in 1984 \cite{RG84}, it has been successfully applied to study excited state properties of a wide range of materials \citep{Carsten}. Compared to other \emph{ab-initio} methods, such as many-body perturbation theory, TDDFT provides a similar level of accuracy, but at far less computational cost. 

The real time evolution of electronic charge and spin densities is calculated using TDDFT by solving the single-particle Kohn-Sham (KS) equations.  The effects of electron-electron interactions come into this non-interacting  KS system via an effective potential, the so called Hartree exchange-correlation
(XC) potential.
Although TDDFT is an exact theory for treating systems  under the influence of strong time-dependent external potentials \cite{EMD16,EKD16,RFSR13,SSS17,KEM17,EMD16,EKD16,DESGS18,SSBM17,KDE15}, it is most commonly applied within the weak perturbation limit. When working in this linear regime, one requires the functional derivative of the XC potential, the so called XC kernel. In a practical TDDFT calculation, an approximation to the XC potential and kernel is required. 

There are many different flavors of XC energy functionals in ground-state DFT, which can be divided into the local density approximation (LDA), generalized gradient approximations (GGAs), meta-GGAs, hybrids and Fock-like approximations, comprising the so called \textit{Jacob's ladder} of approximations, where the level of accuracy increases as we climb from LDA to hybrids. 
The performance of these approximations in static ground-state DFT has been well-studied, however much less is known about their behavior in TDDFT (when combined with the adiabatic approximation). Most of the research so far concerns the optical absorption spectra, and, in particular, the failure of simple XC kernels to predict bound excitons. From these studies, we know the importance of describing the long wavelength limit of the XC kernel correctly in order to obtain reasonable exciton binding energies, leading to a number of new approximations \cite{UY16,SDS12,MSR03,SDS16,RORO02,BSV04,SKR03,SOR03}. However for magnetic excitations, only the ALDA XC kernel has been properly studied, e.g. for calculations of the magnon spectra \cite{BES11,HPOE97,KA00,MFB16,ESDB11}, where for many cases it overestimates magnon energies as compared to experiments. 

In the static DFT case, it is well known that going from LDA to GGA improves many ground state properties  \cite{SCS94}. Hence, in the work presented here, we climb up to the next rung of \textit{Jacobs's Ladder} within TDDFT and ask if including gradient corrections to the XC kernel improves the charge and spin excitation spectra. The paper is organized as follows: Section II gives the basic equations of TDDFT and how they may be used to calculate excitation energies via the linear-response susceptibilities. We will also derive the adiabatic GGA (AGGA) XC kernel for non-collinear spin systems in this section. In Section III, we apply the AGGA kernel to first study the Electron Energy Loss Spectra (EELS) for medium- (diamond) and large- (LiF) bandgap insulators. A comparison is made with experiments and the ALDA kernel. Then a more comprehensive study is made for magnetic excitations of simple bulk ferromagnetic systems Fe, Co and Ni, and Heusler materials. Again we compare with experimental results, as well as previous theoretical works, all of which used the ALDA XC kernel. Finally, in Section IV, we give some concluding remarks on the performance of AGGA.

\section{THEORETICAL FORMULATION}

\begin{figure}[h]
 \includegraphics[width=0.375\textwidth]{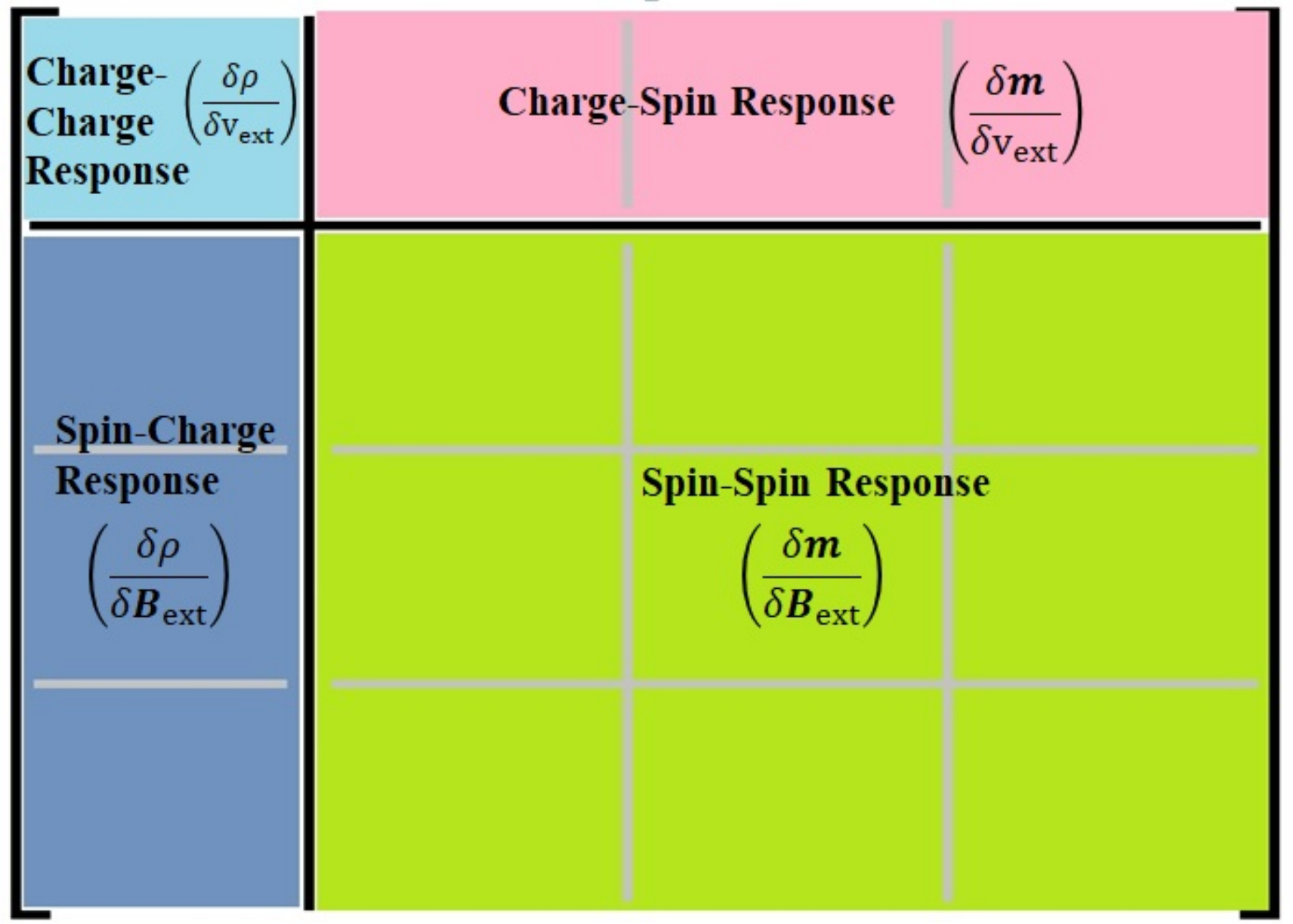}
\caption{\footnotesize{The structure of the fully interacting and non-interacting response functions.}}
\label{f:matrix}
\end{figure}

When the external perturbation is small, the response of the system to this stimulus is studied through a suitable response function. In general, the response of a system to external stimulus can be expanded in a Taylor series with respect to the perturbation. The coefficients of this expansion are the response functions which have useful information embedded in them, such as the optical absorption spectra, Pockels effect, optical rectification, second harmonic generation, Kerr effect etc. In this paper we primarily focus on the first order response functions and particularly on the charge-charge response, ($\delta \rho/\delta v\ext$), and the spin-spin response, ($\delta \textbf{m}/\delta \textbf{B}\ext$), where $\rho,v\ext,\textbf{m},\textbf{B}\ext$ correspond to the charge density, electric scalar potential, magnetization density and magnetic field, respectively. For non-collinear systems, the full response is by:

\begin{equation}
\chi^{\mu\nu} = \dfrac{\delta \rho^{\mu}}{\delta V^{\nu}\ext}
\end{equation}  

where $\rho^{\mu}=[\rho,\textbf{m}]$, $V\ext^{\nu}=[v\ext,\textbf{B}\ext]$, and has a $4\times4$ structure as shown in Fig. (\ref{f:matrix}).

The non-interacting KS linear response functions can easily be calculated in terms of the KS spinors, $\phi(\br)$, using first-order perturbation theory:

\begin{equation}
\label{response}
\begin{split}
\chi_0^{\mu\nu}(\br,\br',\omega) &=\lim_{\eta\to0} \sum_p \sum_q \sigma^\mu \sigma^\nu (f_p-f_q) \\
& \times \dfrac{\phi^*_p(\br)\phi_q(\br)\phi_p(\br')\phi^*_q(\br')}{\omega + (\varepsilon_p-\varepsilon_q) + i\eta} 
\end{split}
\end{equation}

where $f_p,f_q$ denote the occupation number of the $p^{th},q^{th}$ band, respectively, and $\sigma^\mu$ are the four-dimensional counterparts of Pauli spin matrices. 

TDDFT relates this non-interacting response function of the KS system to that of the interacting system via a Dyson-like equation:

\begin{equation}
\label{Dyson}
\begin{split}
\chi^{\mu\nu}(\br,\br',\omega) &= \chi_0^{\mu\nu}(\br, \br',\omega) \\
&+ \sum_{\delta\gamma}\int d^3 r'' \int d^3 r''' \chi_0^{\mu\delta} (\br, \br'',\omega) \Big[ f^{\delta\gamma}_{\text{H}}(\br'', \br''') \\
&+ f\xc^{\delta\gamma}(\br'', \br''',\omega) \Big] \chi^{\gamma\nu}(\br''',\br',\omega)
\end{split}
\end{equation}

where $f_{\text{H}}^{\mu\nu} (\br,\br')=\delta^{\mu0}\delta^{\nu0}v(\br,\br')$ is the Hartree kernel and $v(\br,\br')=1/|\br-\br'|$ is the Coulomb potential, $\omega$ corresponds to frequency and $f\xc^{\mu\nu} (\br, \br',\omega)$ is the XC kernel, which is the Fourier transform of:

\begin{equation}
f\xc^{\mu\nu} (\br,t,\br',t') = \dfrac{\delta \text{V}\xc^{\mu} (\br,t)}{\delta \rho^{\nu} (\br',t')} 
\end{equation} 

where $\text{V}\xc^{\mu}=[v\xc,\textbf{B}\xc]$ is the combined XC potential for the scalar, $v\xc$, and magnetic, $\textbf{B}\xc$, fields.

Since it is convenient to work in reciprocal space for periodic systems, all quantities are represented as matrices via Fourier transformation. The transformed interacting response has the form:

\begin{equation}
\begin{split}
\chi^{\mu\nu}_{\bG,\bG'}& (\bq,\omega) = \\
&\int \int e^{i(\bq+\bG) \cdot \br} \chi^{\mu\nu} (\br,\br',\omega) e^{-i(\bq+\bG') \cdot \br'} d^3r d^3r'
\end{split}
\end{equation}

Here $\bG,\bG'$ are the reciprocal lattice vectors and $\bq$ is the wave vector of the perturbation.  

Conventionally, the excitations are studied in the decoupled limit where the off-diagonal terms of Fig. (\ref{f:matrix}) are set to zero (i.e. $\delta {\bf m}/\delta v\ext=\delta \rho/\delta {\bf B}\ext$=0). This allows us to separate the dielectric response and magnetic response.

Experimental observables may then be extracted from the response functions, for example, the inverse dielectric function is  

\begin{equation}
\epsilon^{-1} = 1 + v \Big( \dfrac{\delta \rho}{\delta v} \Big)
\end{equation}

The imaginary part of $\epsilon^{-1}$ gives the EELS whereas the imaginary part of $\epsilon$ gives the absorption spectrum. Likewise, the neutron scattering cross-section is proportional to the transverse magnetic response \cite{LM81}

\begin{equation}
\begin{split}
\dfrac{d^2 \sigma}{d\Omega d\omega} &\propto \Big\{ (1-\kappa^2_z) Im [\chi^{zz} (\bq,\omega)] \\
&+ \dfrac{1}{4}(1-\kappa^2_z) Im [\chi^{-+} (\bq,\omega) + \chi^{+-} (\bq,\omega)] \Big\}
\end{split}
\end{equation}

where $\kappa_z=(\textbf{k}_f-\textbf{k}_i)_z/|\textbf{k}_f-\textbf{k}_i|$ is related to the $\bq$-vector through $\bq=\textbf{k}_f-\textbf{k}_i$ and is folded back into the first Brillouin Zone (BZ). Here, the transverse terms comprise $\chi^{-+}(=2\chi^{xx}+2i\chi^{xy})$ and $\chi^{+-}(=2\chi^{xx}-2i\chi^{xy})$. The term $\chi^{zz}$ does not contribute to the spin-flip excitations, rather it's the transverse terms of the magnetic susceptibility which give rise to the Stoner and magnon excitations. 

\subsection{GGA Kernel}

Interactions between the electrons are encapsulated in the XC potential. Knowledge of the exact form of $v\xc$, and hence $f\xc$, would lead to the exact solution of all many-body problems. However, the exact form of $v\xc$ is not known and approximations are required for all practical calculations. Here we will derive the XC kernel for GGA functionals within the adiabatic approximation (AA), which is semi-local in space and local in time. For the spin unpolarized case, the XC energy functional, $E\xc$, depends not only on the density, $n(\br)$, but also on its gradient, ${\mathbf\nabla} n(\br)$, at each point $\br$ in space. The XC potential and kernel can be obtained from first and second order functional derivatives, respectively, of $E\xc$ with respect to density i.e.,

\begin{equation}
v\xc[\rho](\br)=\frac{\delta E\xc[\rho]}{\delta \rho(\br)}
\end{equation} 

\begin{equation}
f\xc[\rho](\br,\br')=\frac{\delta^2 E\xc[\rho]}{\delta \rho(\br) \delta \rho(\br')}
\end{equation}

where $E\xc[\rho]=\int e\xc(\rho,\bn \rho) (\vec{r}) d^3r$ for the GGA functional and $e\xc$ is the XC energy density.

The variation of the XC energy is defined by: 

\begin{equation}
\label{eq3}
\begin{split}
\delta E\xc&= E\xc[\rho+\delta \rho]-E\xc[\rho] \\
&= \int v\xc[\rho] (\br) \delta \rho(\br) d^3r \\
&+\dfrac{1}{2} \int \int f\xc[\rho](\br,\br') \delta \rho(\br) \delta \rho(\br') d^3r d^3r' + \cdots
\end{split}
\end{equation}

Taylor expanding the energy density up to first order gives

\begin{equation}
\begin{split}
e\xc(\rho+\delta \rho,\bn \rho + \bn \delta \rho) (\br) &= e\xc(\rho,\bn \rho) (\br)\\
&+ \frac{\partial e\xc(\rho,\bn \rho)}{\partial \rho}  (\br) \delta \rho(\br) \\
&+ \frac{\partial e\xc(\rho,\bn \rho)}{\partial \bn \rho} (\br) \cdot \bn \delta \rho(\br) \\
\end{split}
\end{equation}

leading to the expansion of energy functional,

\begin{equation}
\begin{split}
E\xc[\rho+\delta \rho, \bn \rho + \bn \delta \rho] &= E\xc[\rho,\bn \rho] \\
&+ \int d^3r \frac{\partial e\xc(\rho,\bn \rho)}{\partial \rho} (\br) \delta \rho(\br) \\
&+ \int d^3r \frac{\partial e\xc(\rho,\bn \rho)}{\partial \bn \rho} (\br) \cdot \bn \delta \rho(\br) \\
\end{split}
\end{equation}

and

\begin{equation}
\begin{split}
\delta E\xc =\int d^3r \Big[ & \frac{\partial e\xc(\rho,\bn \rho)}{\partial \rho} (\br) \delta \rho(\br) \\
&+ \frac{\partial e\xc(\rho,\bn \rho)}{\partial \bn \rho} (\br) \cdot \bn \delta \rho(\br) \Big] 
\end{split}
\end{equation}

Carrying out integration by parts of the second term gives us:

\begin{equation}
\begin{split}
\label{eq4}
\delta E\xc = \int d^3r \Big[ & \frac{\partial e\xc(\rho,\bn \rho)}{\partial \rho} (\br) \\
&- \Big\{ \bn \cdot \frac{\partial e\xc(\rho,\bn \rho)}{\partial \bn \rho} (\br) \Big\} \Big] \delta \rho(\br)
\end{split}
\end{equation}

Comparing Eq. \eqref{eq4} with Eq. \eqref{eq3}, we find the XC potential as:
 
\begin{equation}
\begin{split}
v\xc[\rho](\br)&= \frac{\partial e\xc(\rho,\bn \rho)}{\partial \rho} (\br) - \Big\{ \bn \cdot \frac{\partial e\xc(\rho,\bn \rho)}{\partial \bn \rho} (\br) \Big\} \\
&= \frac{\partial e\xc(\rho,\bn \rho)}{\partial \rho} (\br) \\
&- 2 \Big\{ \bn \cdot \Big( \frac{\partial e\xc(\rho,\bn \rho)}{\partial \sigma} (\br) \bn \rho \Big) \Big\}
\end{split}
\end{equation}

where $\sigma=\bn \rho \cdot \bn \rho$ is often used in practice. Variation of this potential to first order will give the kernel:

\begin{equation}
\begin{split}
\delta v\xc(\br)&=v\xc[\rho+\delta \rho,\bn \rho+ \bn  \delta \rho] (\br) - v\xc[\rho,\bn \rho](\br) \\
&= \frac{\partial^2 e\xc}{\partial \rho^2} (\br) \delta \rho(\br) + \frac{\partial^2 e\xc}{\partial \bn_j \rho \partial \rho} (\br)  \bn_j \delta \rho(\br) \\
&- \bn_k  \Big[ \frac{\partial^2 e\xc}{\partial \rho \partial \bn_k \rho} (\br) \delta \rho(\br) \\
&+ \frac{\partial^2 e\xc}{\partial \bn_j \rho \partial \bn_k \rho} (\br) \bn_j \delta \rho(\br) \Big] \\
\end{split}
\end{equation}

Integrating these terms individually by introducing a delta function gives us the kernel for GGA functional \cite{NV11,EB04,DBR94}

\begin{equation}
\label{fxc}
\begin{split}
f\xc(\br,\br') &= \delta(\br-\br')  \Big[  \frac{\partial^2 e\xc}{\partial \rho \partial \rho} (\br') 
- \bn^{r'}_j \frac{\partial^2 e\xc}{\partial \bn_j \rho \partial \rho} (\br') \Big] \\
&- \bn^{r'}_j \Big[ [ \bn^{r'}_k  \delta(\br-\br') ] \frac{\partial^2 e\xc}{\partial \bn_j \rho \partial \bn_k \rho} (\br') \Big]
\end{split}
\end{equation}

Repeating the above derivation for the spin polarized case (see Appendix A) gives us two equations comprising the symmetric terms $f^{\alpha\alpha}\xc(\br,\br')$, $f^{\beta\beta}\xc(\br,\br')$ and the asymmetric terms $f^{\alpha\beta}\xc(\br,\br')$, $f^{\beta\alpha}\xc(\br,\br')$ of the XC kernel matrix, where $\alpha$ and $\beta$ label the up and down spins, respectively,

\begin{equation}
\label{faa}
\begin{split}
f^{\alpha\alpha}\xc(\br,\br') &= \delta(\br-\br') \Big[ \frac{\partial^2 e\xc}{\partial \rho_\alpha \partial \rho_\alpha} (\br')
- \bn^{r'}_k \frac{\partial^2 e\xc}{\partial \bn_k \rho_\alpha \partial \rho_\alpha} (\br') \Big] \\
&- \bn_k^{\br'} \Big[ [\bn_j^{\br'} \delta(\br-\br')]  \Big( \frac{\partial^2 e\xc}{\partial \bn_k \rho_\alpha \partial \bn_j \rho_\alpha} (\br') \Big) \Big]  
\end{split}
\end{equation}

\begin{equation}
\label{fab}
\begin{split}
f^{\alpha\beta}\xc(\br,\br') &= \delta(\br-\br') \Big\{ \frac{\partial^2 e\xc}{\partial \rho_\beta \partial \rho_\alpha} (\br') \Big\} \\
&- \Big\{ [\bn^{r'}_j \delta(\br-\br')]  \Big( \frac{\partial^2 e\xc}{\partial \rho_\alpha \partial \bn_j \rho_\beta }  (\br) \Big) \Big\} \\
&+ \Big\{ [\bn^{r'}_j \delta(\br-\br')]  \Big( \frac{\partial^2 e\xc}{\partial \rho_\beta \partial \bn_j \rho_\alpha}  (\br') \Big) \Big\} \\
&- \bn^{r'}_k \Big\{ [\bn^{r'}_j \delta(\br-\br')]  \Big( \frac{\partial^2 e\xc}{\partial \bn_k \rho_\beta \partial \bn_j \rho_\alpha} (\br') \Big) \Big\} 
\end{split} 
\end{equation}

which is extended to non-collinear systems via the K\"ubler method \cite{KHSW88}, see Eqs. (\ref{f00})-(\ref{fij}).

\section{COMPUTATIONAL DETAILS}

\begin{figure}[b]
\begin{subfigure}{0.5\textwidth}
  \centering
  \includegraphics[width=0.75\textwidth,angle=-90]{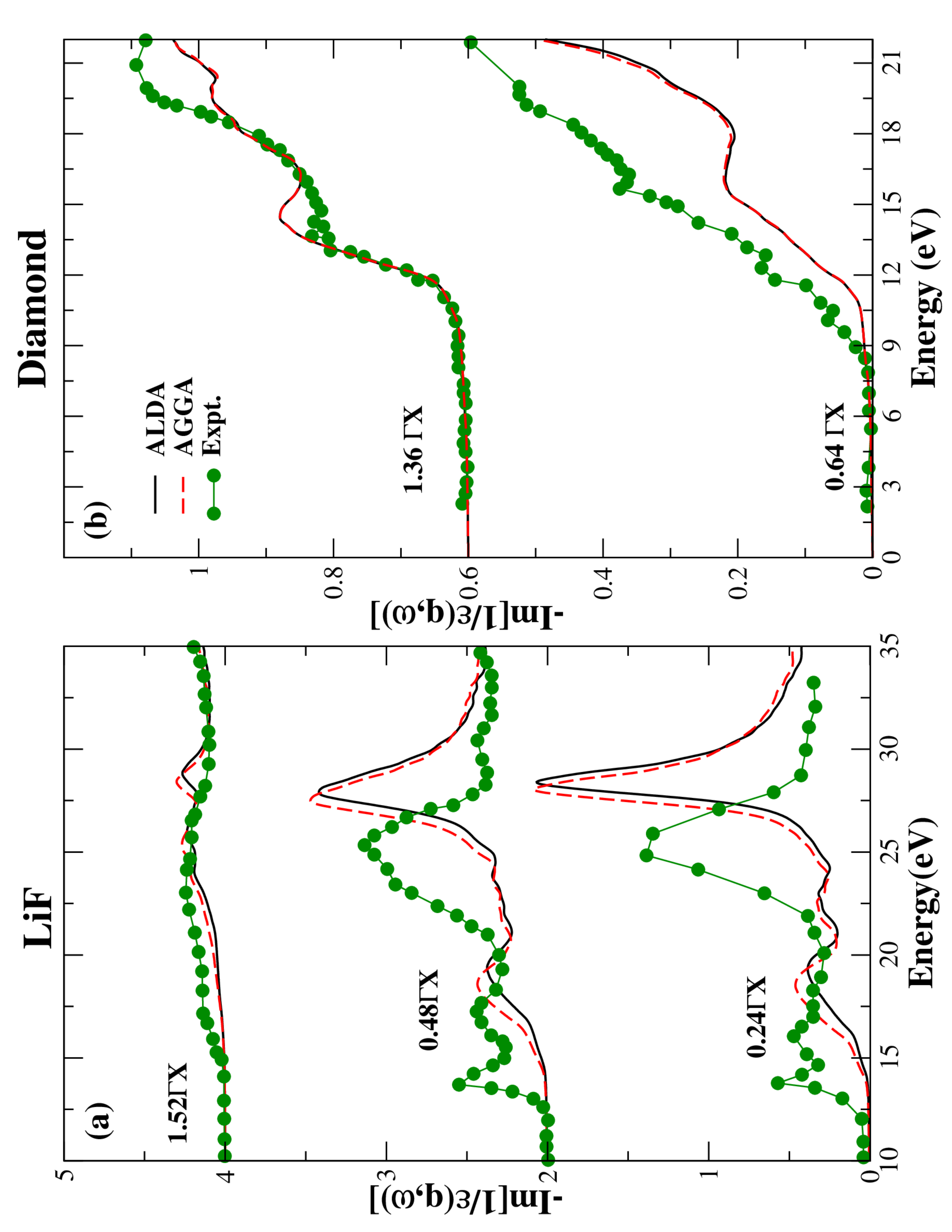}
 \end{subfigure} \\
\caption{\footnotesize{Electron energy loss spectra given by imaginary part of the inverse dielectric tensor for different values of \textbf{q} (indicated in the figure) as a function of photon energy for (a) LiF and (b) Diamond, using the AGGA kernel (red dashed), the ALDA kernel (black line) and the experimental data \cite{CSS00} (green dots).}}
\label{Optical}
\end{figure}

\begin{figure*}[t]
 \includegraphics[width=0.4\textwidth,angle=-90]{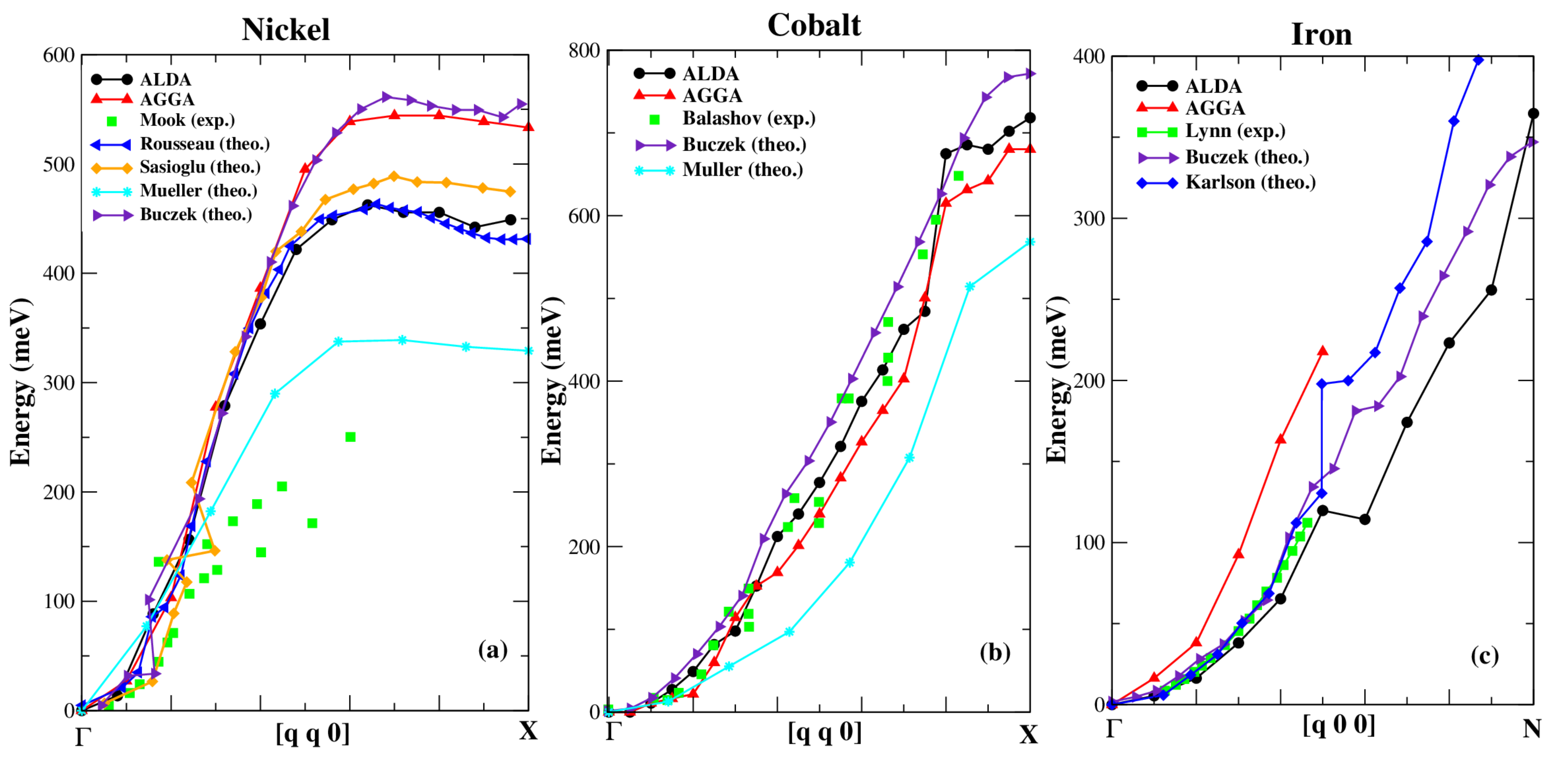}
\caption{\footnotesize{Magnon dispersion spectrum for (a) fcc Nickel, (b) fcc Cobalt along the $\Gamma$X direction and (c) bcc Iron along the $\Gamma$N direction calculated using the ALDA kernel (black dots) and AGGA kernel (red triangles). A comparison is made with reported theoretical work \cite{BES11,REB12,SSF10,MFB16,KA00} which used the ALDA kernel only and also the experimental result (green squares) taken from Mook et al. \cite{LM81,MP85} for Nickel, Balashov et al. \cite{Balashov} for Cobalt, and Lynn \cite{Lynn75} for Iron.}}
\label{f:NiCoFe}
\end{figure*}

All calculations are performed using the all-electron full-potential linearized augmented plane wave electronic structure code ELK \cite{elk} with PW91 (LDA) \cite{PW92} and PBE (GGA) \cite{PBE96} functionals. For diamond and LiF, a fcc crystal structure with lattice spacing of $3.5599$  \text{\AA} and $4.0259$ \text{\AA}, respectively, is used. A dense k-point grid is required to obtain the response functions, hence the BZ is sampled on a k-point grid of $25\times 25\times 25$ for both. The method to obtain response functions is a two-step procedure, firstly a ground-state calculation is done to obtain the converged density and potentials. The scissor operator has been used to correct the optical band-gap by 1.306 eV and 5.06 eV for diamond and LiF, respectively. Then the EELS spectra of LiF and diamond are obtained from the LDA and GGA kernels using the corrected band gaps.

The magnon spectra are highly sensitive to a number of parameters, hence convergence has to be checked with respect to k-point grid and the number of $\bG$-vectors. In these calculations we have used a $40 \times 40 \times 40$ k-point grid. The response functions are expanded in $\bG$-space with the length of $\bG$-vector up to 6 Bohr$^{-1}$. The lattice constant used for Co$_2$MnSi is $5.640$ \text{\AA} and for NiMnSb is $5.897$ \text{\AA}. The spectra are corrected by choosing convergence parameters so that it best satisfies Goldstone's theorem.

\section{RESULTS}

\subsection{Semiconductor Spectra}

It is well known that the $\bq$-dependent behavior of the XC kernel is of vital importance for predicting the optical response of materials. For example, the XC kernel must go as $1/\bq^2$ in order to capture excitonic effects \cite{RORO02,SOR03,SDG14,SDSMG15,ANSS16,SDSG11} in the long-wavelength limit ($\bq\to 0$). However, the first rung on \textit{Jacob's ladder}, the ALDA, does not display any $\bq$-dependence, owing to the local approximation for the XC energy. This explains why ALDA does not yield excitonic peaks \cite{SDG14}. The second rung consists of semi-local functionals which include information not just about the density but also its gradients. In this case, it has been shown that the AGGA kernel shows $\bq^2$ behaviour \cite{GGG97}. Hence we explore if this has an impact on the spectra.

In Fig. \ref{Optical}(a) we plot the EELS for (i) LiF, which is a large bandgap material with a bound exciton, and (ii) diamond which is a medium bandgap material with excitonic effects appearing as a shift in the spectral weight towards lower energies. For LiF, we can see that AGGA shifts the peak energies for $\bq=0.24{\bf \Gamma X}, 0.48 {\bf \Gamma X}$ towards the correct experimental values and also makes the peaks more pronounced, as compared to ALDA. Both ALDA and AGGA fail to capture the excitonic peak at $13$ eV as neither has the correct $1/\bq^2$ behavior in the long-wavelength limit. Outside the first BZ ($\bq =1.52 {\bf \Gamma X}$), ALDA and AGGA exhibit similar behaviour. For diamond, neither AGGA nor ALDA captures the shift in spectral weight as can be seen in Fig. \ref{Optical}(b). In fact there is little difference between the results obtained using the two approximations.

To conclude this section, the $\bq$-dependence of the AGGA kernel slightly improves upon the ALDA results, although neither captures the excitonic effects. 

\begin{figure*}[t]
\begin{tabular}{ c c c c c }
        &  & ALDA & & AGGA \\
       \parbox[t]{3mm}{\multirow{-18}{*}{\rotatebox[origin=c]{90}{Nickel}}} & 
       \parbox[t]{3mm}{\multirow{-24}{*}{(a)}} & 
       \includegraphics[width=0.4\textwidth]{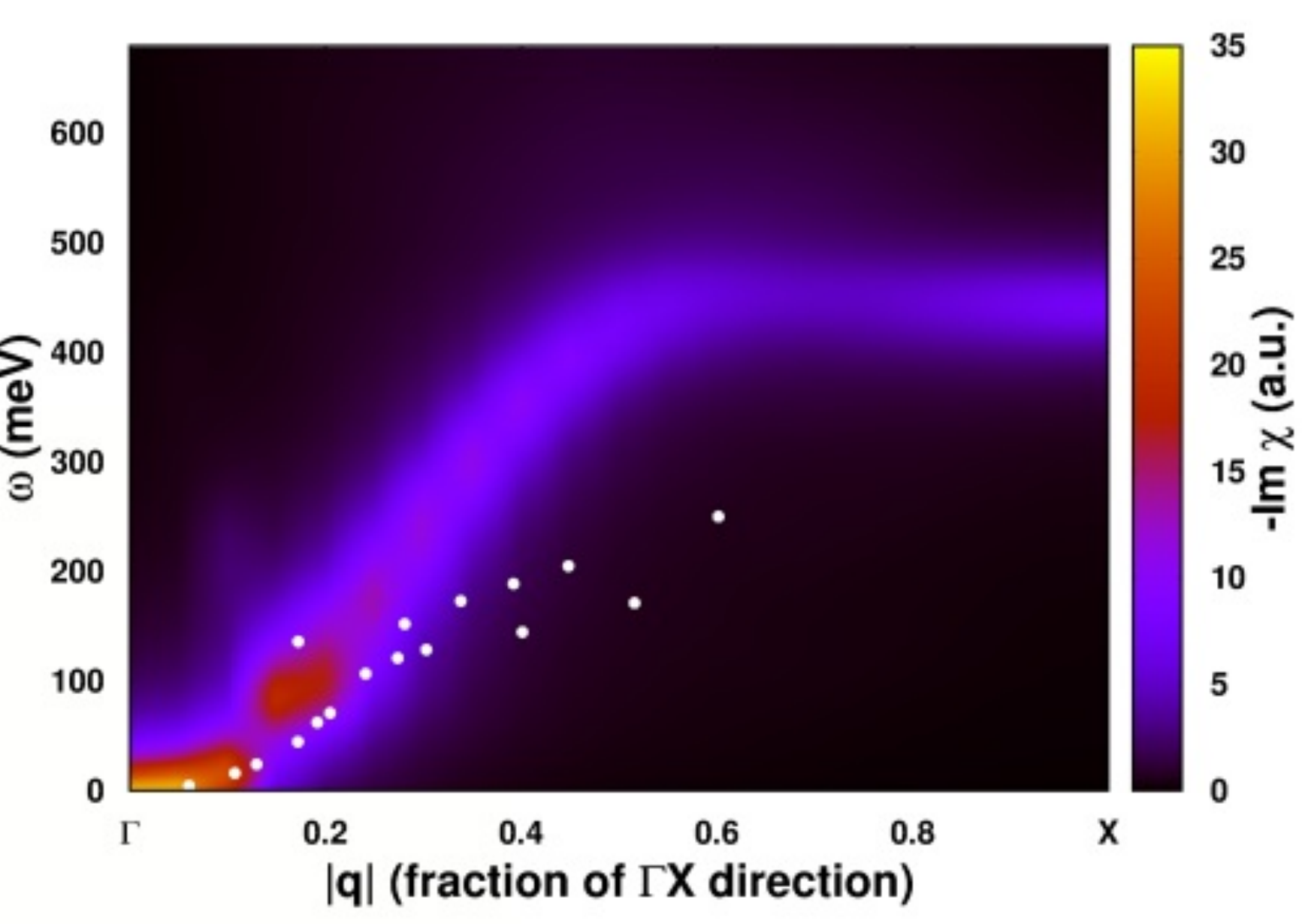} &
       \parbox[t]{3mm}{\multirow{-24}{*}{(b)}} & 
      \includegraphics[width=0.4\textwidth]{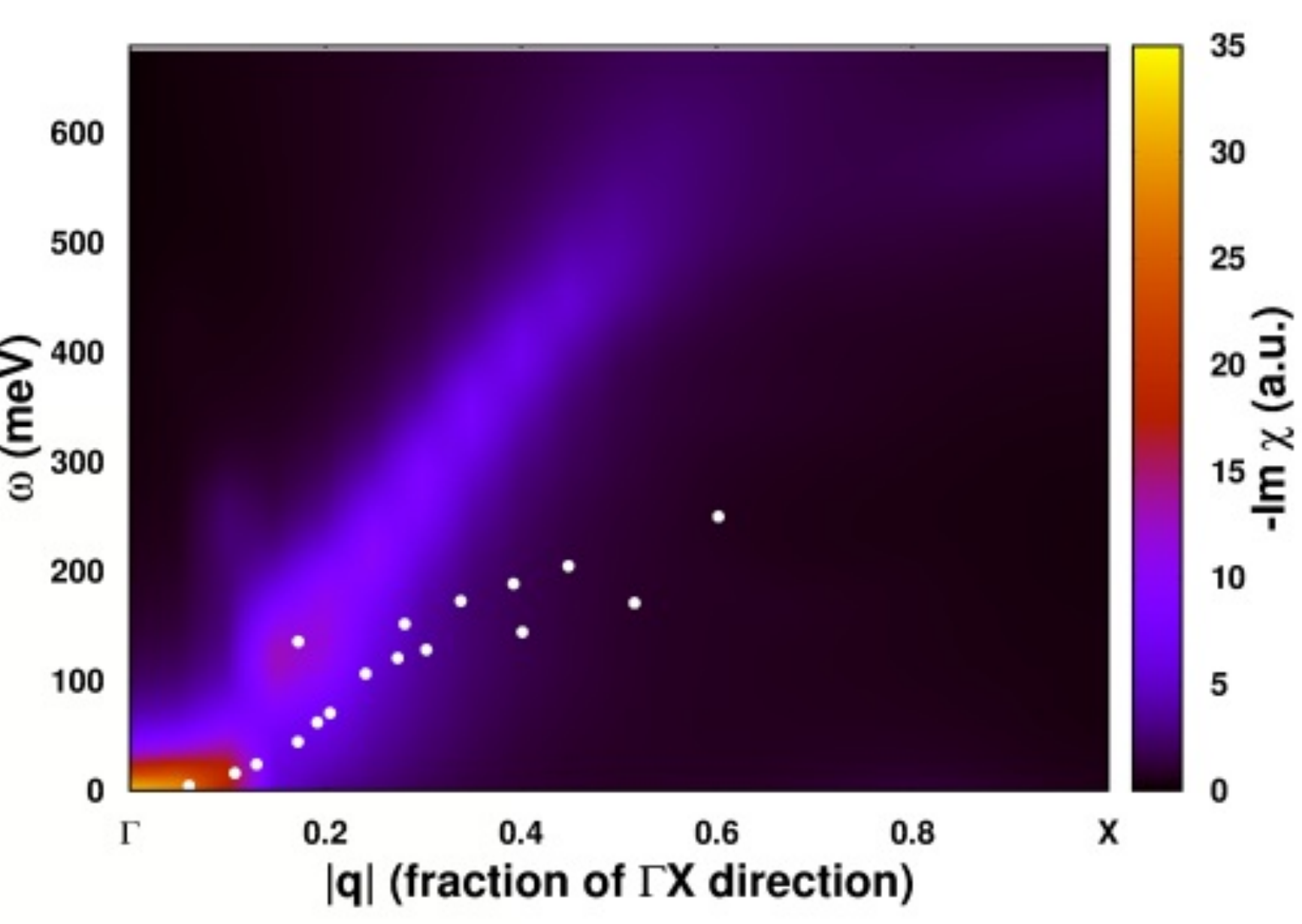}  \\
       \parbox[t]{1mm}{\multirow{-18}{*}{\rotatebox[origin=c]{90}{Cobalt}}} & 
       \parbox[t]{1mm}{\multirow{-24}{*}{(c)}} &
      \includegraphics[width=0.4\textwidth]{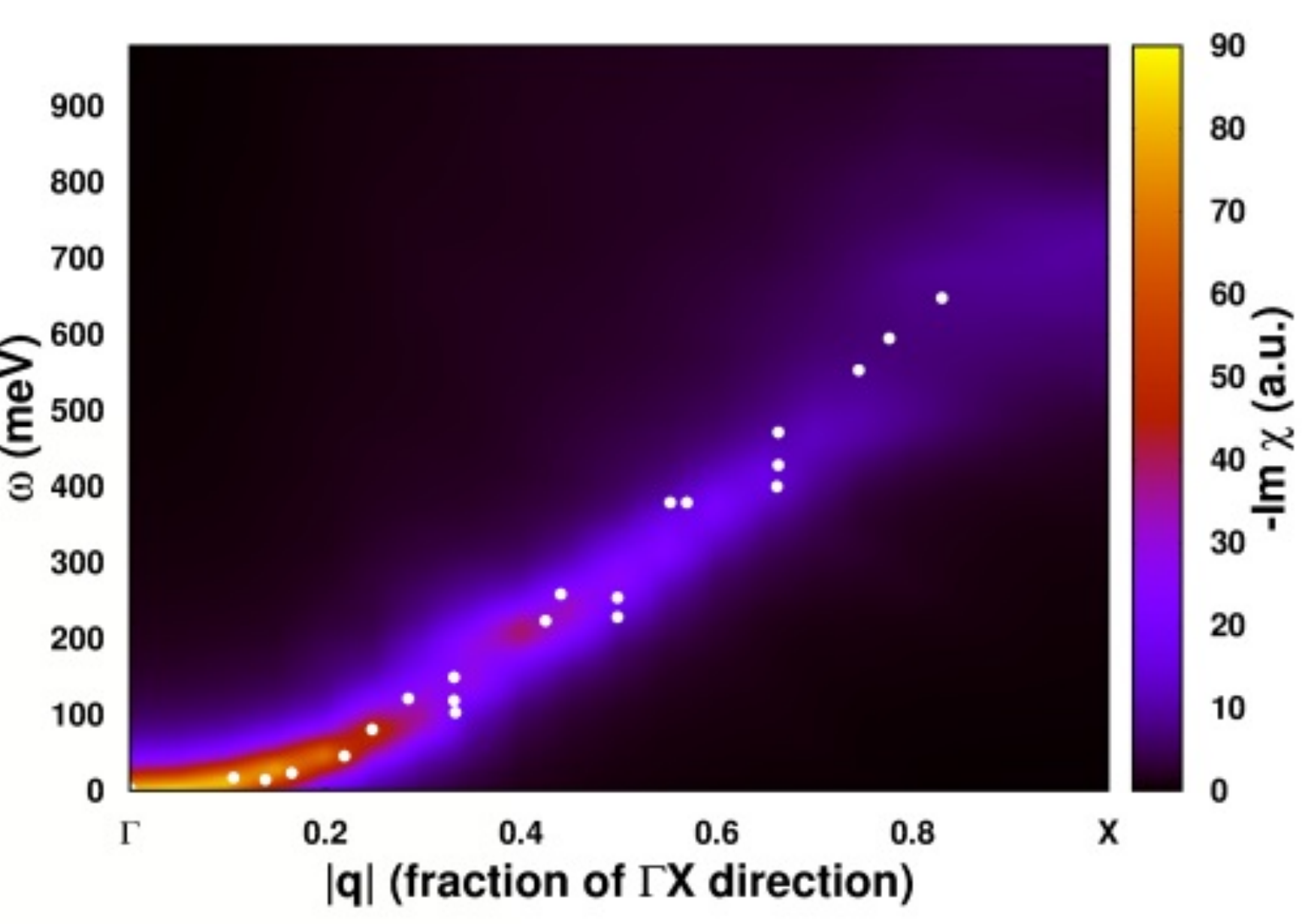} &
      \parbox[t]{3mm}{\multirow{-24}{*}{(d)}} & 
      \includegraphics[width=0.4\textwidth]{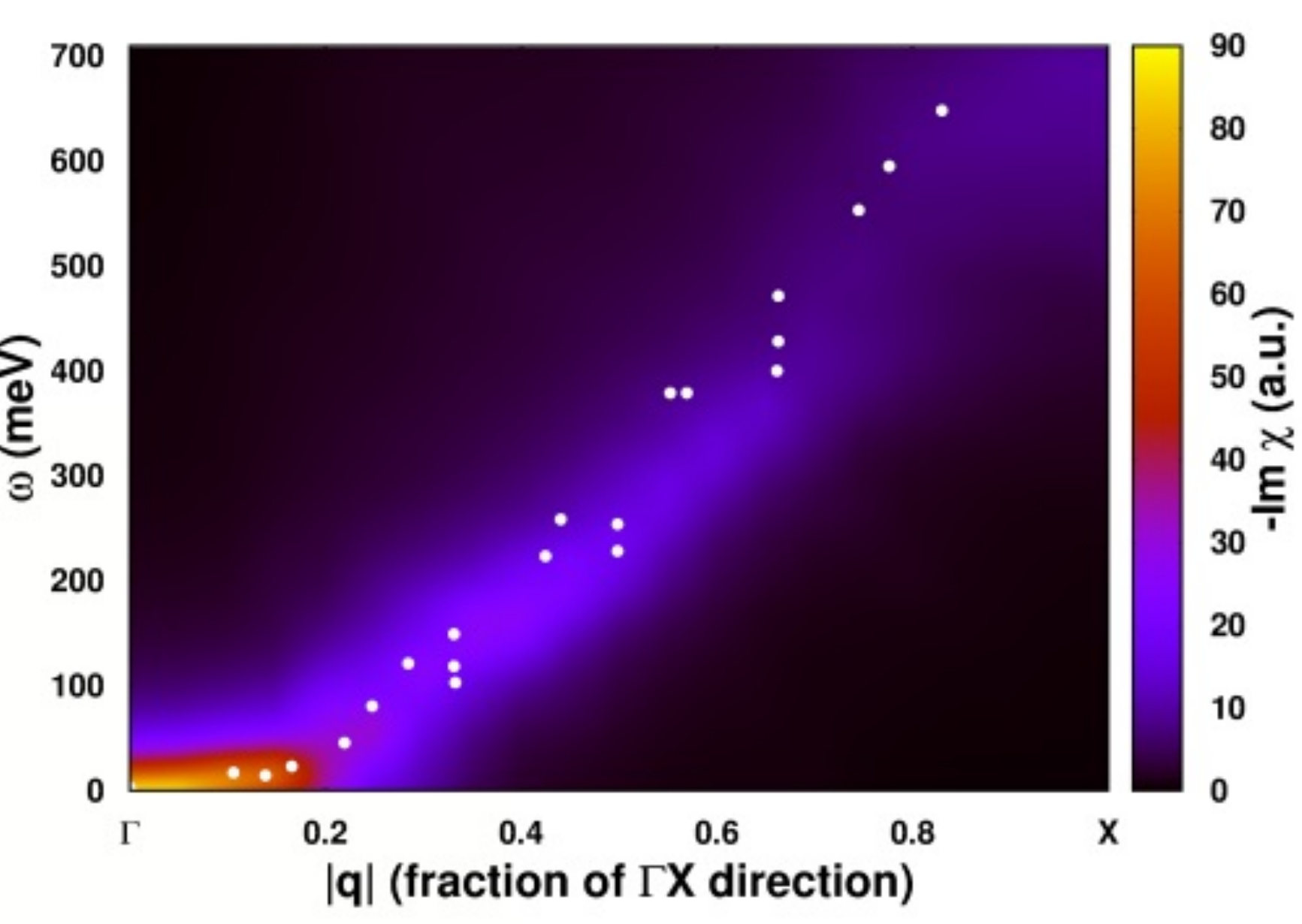}  \\
      \parbox[t]{3mm}{\multirow{-18}{*}{\rotatebox[origin=c]{90}{Iron}}} &
      \parbox[t]{1mm}{\multirow{-24}{*}{(e)}} 
      & \includegraphics[width=0.4\textwidth]{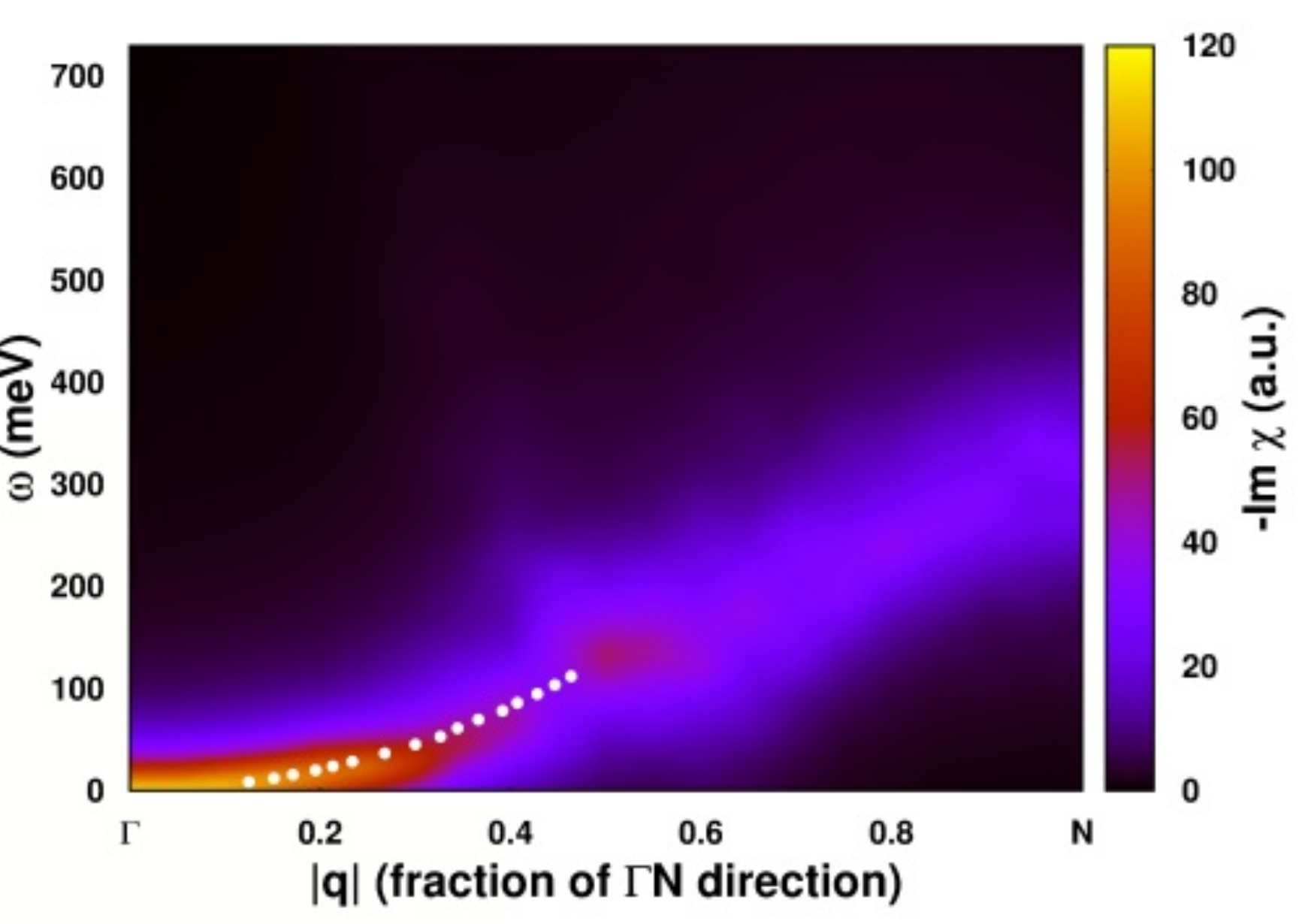} &
      \parbox[t]{3mm}{\multirow{-24}{*}{(f)}} & 
      \includegraphics[width=0.4\textwidth]{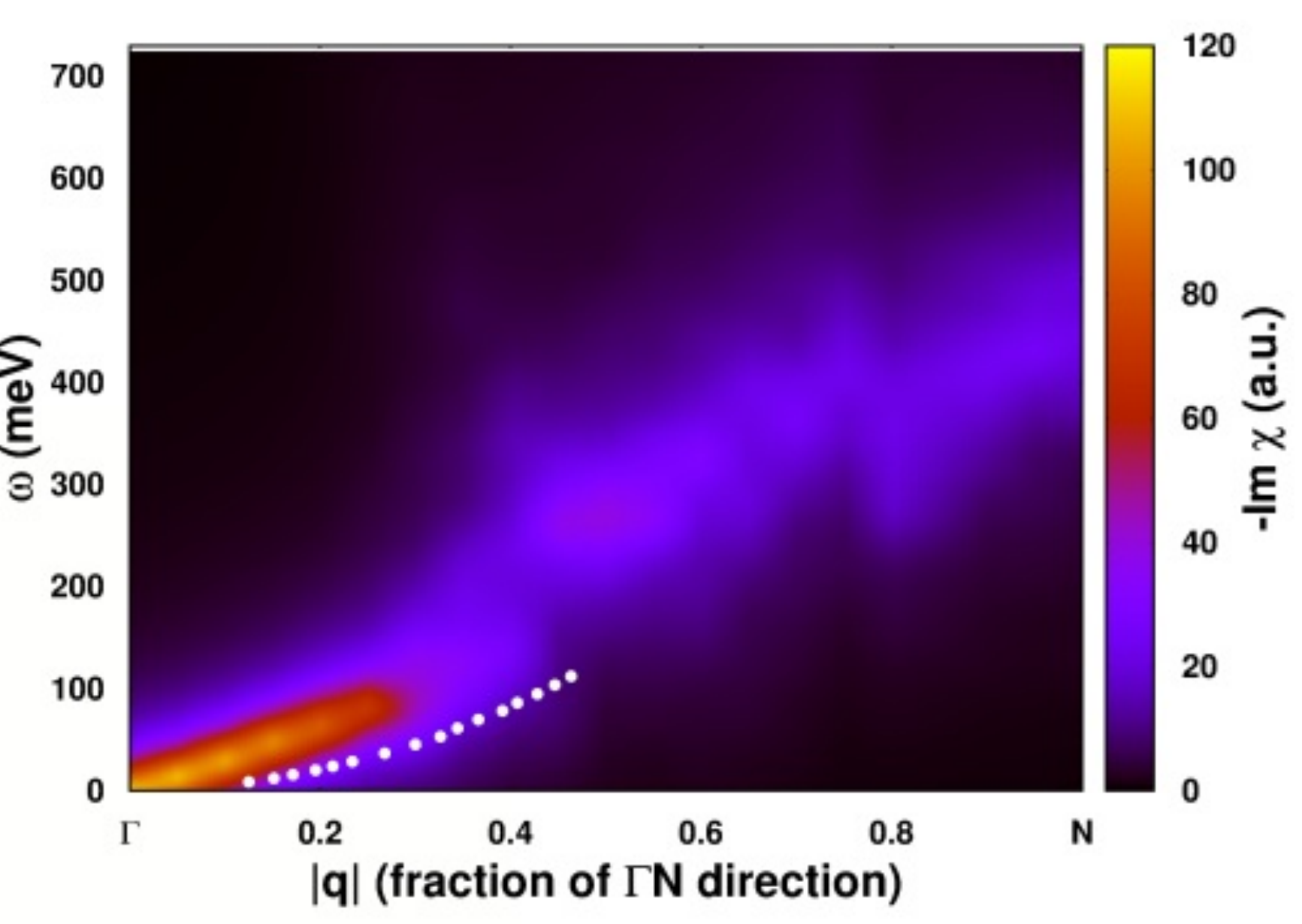}  \\
      \end{tabular}
\caption{\footnotesize{The imaginary part of the interacting response of Nickel, Cobalt, and Iron using the ALDA kernel (a,c,e) and the AGGA kernel (b,d,f), and the experimental results \cite{MP85,Balashov,Lynn75} (white dots).}} 
\label{f:NiCoFeChiT}
\end{figure*}

\subsection{Spin Excitation Spectra}

For a computational method to be fully \textit{ab-initio}, given the atomic composition, it should first predict the equilibrium geometry. Having found the minimum energy crystal structure from ground-state calculations, we can then calculate the excited state properties, all without reference to experimental data. Only those methods which follow this prescription can be considered fully predictive.

We begin by reviewing ground-state DFT calculations with experimental and optimized lattice parameters. The results are summarized in Table \ref{t:lat}. From this we conclude that (i) GGA is very good in reproducing the structures of materials whereas (ii) LDA is better in predicting the magnetic moments.

\begin{table}[h]
\caption{\footnotesize{Equilibrium lattice parameters, $a_0$ (in \AA), calculated using the 3$^{rd}$ order Birch-Murnaghan equation of state. Magnetic moments, $m$ (in $\mu_B$), obtained at the equilibrium and the experimental lattice parameters}.}
\begin{tabular}{c c c c c c c }
\hline \hline
 & $a_0$ (exp.) & $a_0$ (LDA) & $a_0$ (GGA) & $m_{exp.}$  &  $m_{LDA}$  & $m_{GGA}$ \\
\hline
Ni(fcc) & 3.524\footnote{\label{1st}Reference [\onlinecite{VJC07}].}    & 3.436  & 3.527 &  0.60\footref{1st} &  0.591 & 0.636 \\
Co(fcc) & 3.539\footnote{\label{2nd}Reference [\onlinecite{SpringerB1}].}     & 3.429  & 3.525 &  1.52\footref{2nd} &  1.525 & 1.641 \\
Fe(bcc) & 2.8665\footnote{\label{3rd}Reference [\onlinecite{SpringerB2}].}    & 2.743  & 2.836 &  2.08\footref{3rd} &  1.996 & 2.174  \\
\hline \hline
\end{tabular}
\label{t:lat}
\end{table}

Within TDDFT the magnon spectra of a system can be calculated from the transverse response function $\chi^{-+}(\bq,\omega)$, which is found using the Dyson-like equation, Eq. (\ref{Dyson}). The excitations in $\chi^{-+}(\bq,\omega)$ originate from two sources, (1) renormalized poles of the KS response $\chi_0$ corresponding to the Stoner continuum of single-particle spin-flips and (2) additional peaks created by the XC kernel corresponding to spin-wave excitations.

\begin{figure*}[t]
 \includegraphics[width=0.4\textwidth,angle=-90]{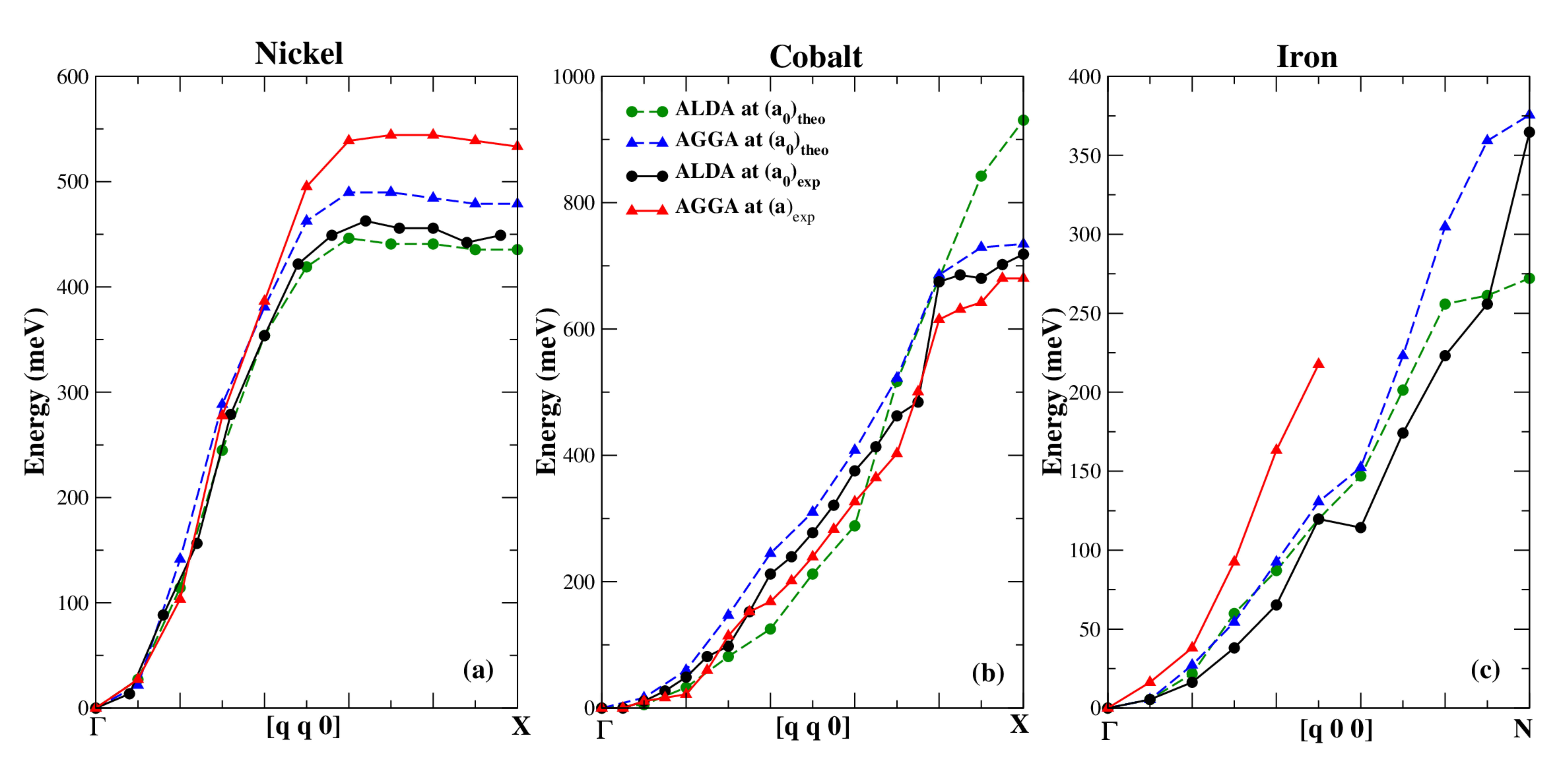}
\caption{\footnotesize{The magnon spectrum with the theoretical and experimental lattice parameters for (a) fcc Nickel, (b) fcc Cobalt along the $\Gamma$X direction and (c) bcc Iron along the $\Gamma$N direction  calculated using the ALDA kernel (dots) and AGGA kernel (triangles). The lattice parameters are given in Table \ref{t:lat}.}}
\label{f:NiCoFea0}
\end{figure*}

To find the magnon dispersion, we calculate Im$\{\chi^{-+}(\bq,\omega)\}$ for each $\bq$ value, extract the magnon peak position and then plot these as a function of $\bq$. This is shown in Fig. (\ref{f:NiCoFe}) for nickel, cobalt, and iron along with experimental measurements and ALDA calculations reported in the literature \cite{BES11,SSF10,MFB16,S98,HPOE97,KA00,MFB16}. We first note that the ALDA results reported in this paper are consistent with previously reported results. In the following, we will discuss each material individually before commenting on the general behavior of the AGGA XC kernel.

\begin{figure}[b]
  \includegraphics[width=0.375\textwidth,angle=-90]{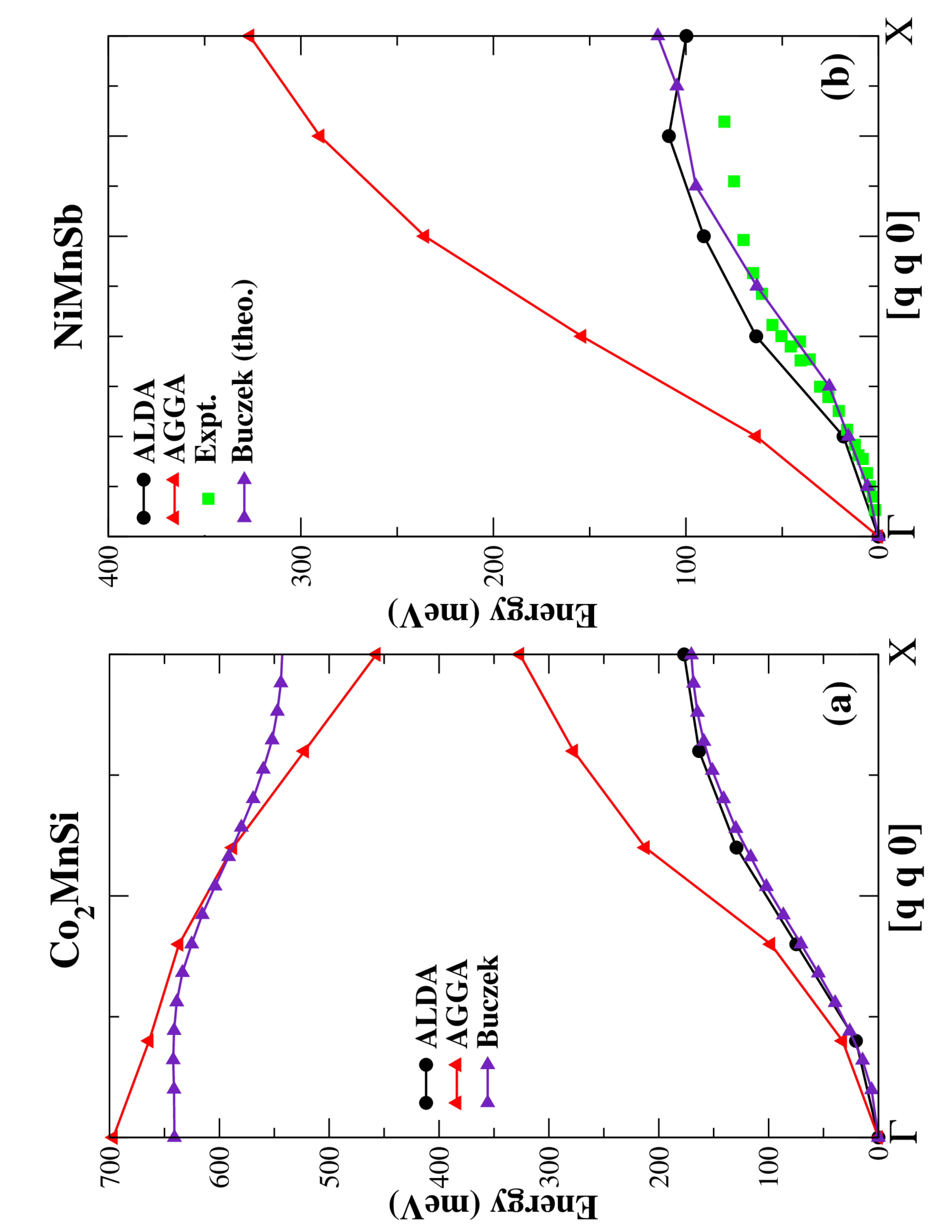}
\caption{\footnotesize{(a) The magnon spectra of Co$_2$MnSi using the ALDA (black dots) and AGGA (red triangles) kernels and compared with Buczek calculations \cite{BEBS09} with the ALDA kernel (violet triangle left). (b) The magnon spectra of NiMnSb using ALDA and AGGA kernel and the experimental results \cite{HP97} (green squares).}}
\label{f:Heuslers}
\end{figure}

\begin{figure*}[t]
\begin{tabular}{c c c c}
    \parbox[t]{3mm}{\multirow{-24}{*}{(a)}} & 
   \includegraphics[width=0.375\textwidth]{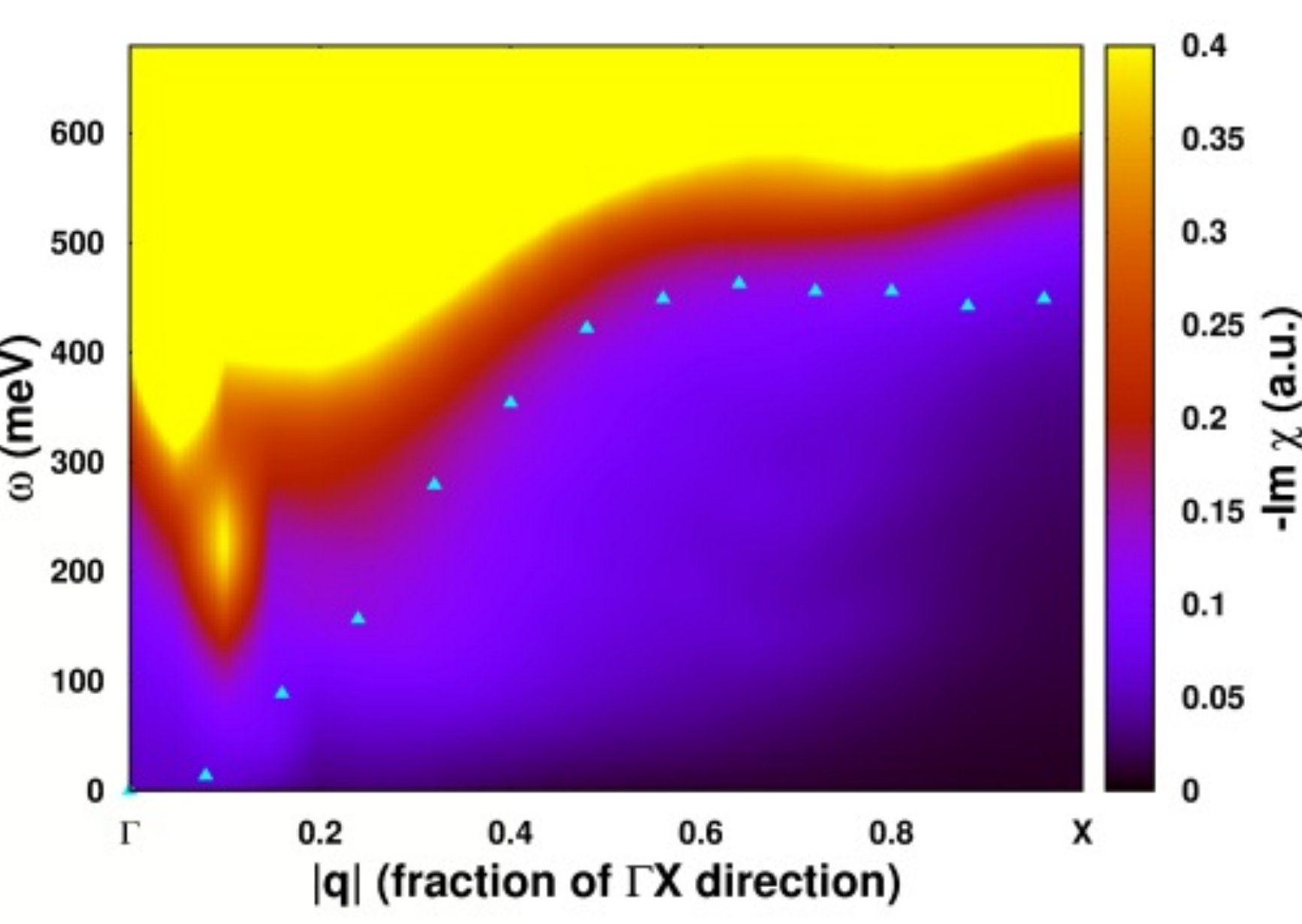} &
    \parbox[t]{3mm}{\multirow{-24}{*}{(b)}} & 
    \includegraphics[width=0.375\textwidth]{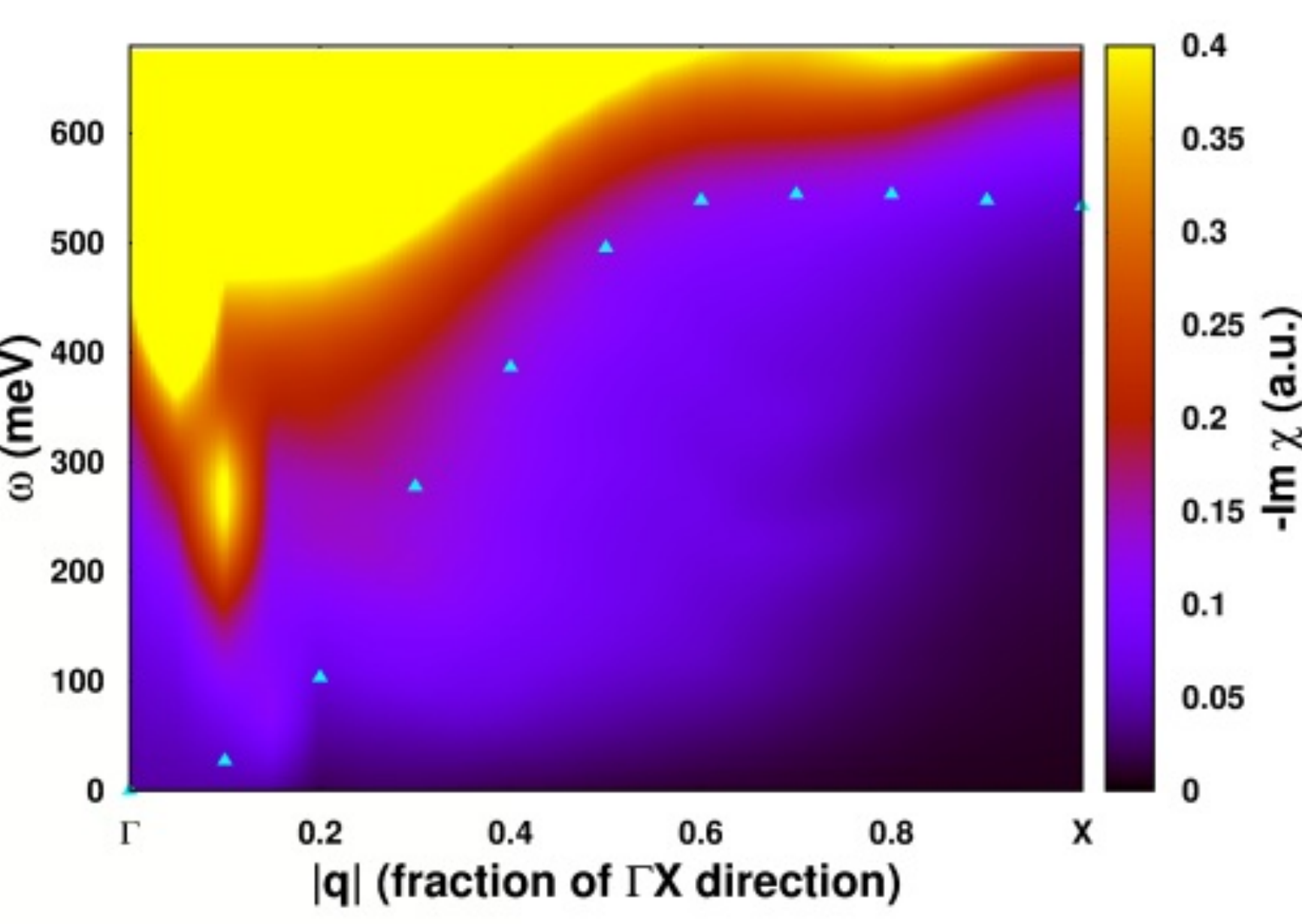} \\
\end{tabular}
\caption{\footnotesize{Imaginary part of the non-interacting response function for Nickel using (a) LDA  and (b) GGA and also the corresponding theoretical magnon spectra for comparison (cyan triangles).}}.
\label{f:NiChi0T}
\end{figure*}

For Ni, Fig. \ref{f:NiCoFe}(a), both ALDA and AGGA show quantitatively the same behaviour from the BZ center to $|\bq|=0.4$. As we move further  away from the zone center, the AGGA kernel tends to deviate from ALDA until it becomes $\approx 80$ meV higher in energy at the zone boundary. For Co, Fig. \ref{f:NiCoFe}(b), the experimental results are well captured by both ALDA and AGGA calculations. For Fe, Fig. \ref{f:NiCoFe}(c), in contrast to ALDA which reproduces the experimental values, the AGGA dispersion overestimates the magnon energies. Beyond the zone center, the transverse response function obtained using AGGA becomes too broad to assign a single energy to the excitation peaks.

The strength and width of the peak in Im$\{\chi^{-+}(\bq,\omega)\}$ is related to the scattering amplitude and lifetime of the magnon, respectively. To visualize how these properties change throughout the BZ, we can make a 2D contour plot of Im$\{\chi^{-+}(\bq,\omega)\}$. These are shown for both ALDA and AGGA in Fig. (\ref{f:NiCoFeChiT}) for Ni, Co and Fe.

Beginning again with Ni, we see that the peaks in Im$\{\chi^{-+}(\bq,\omega)\}$ obtained by using ALDA (Fig. \ref{f:NiCoFeChiT}(a)) are stronger in intensity and better resolved than AGGA (Fig. \ref{f:NiCoFeChiT}(b)). There exists a high probability of observing a magnon at the BZ boundary with ALDA whereas it is suppressed significantly beyond $\bq=0.5 {\bf \Gamma X}$ with AGGA. We also observe a strong suppression in the magnon intensity between $|\bq|=0.1$ and $|\bq|=0.2$ for AGGA and ALDA (see Figs. \ref{f:NiCoFeChiT}(a) and (b)). Experimentally, Paul et al. \cite{MP85} measured a disruption to the magnon dispersion at $|\bq|=0.2$, where they observed a split into optical and acoustic branches. While neither ALDA nor AGGA shows two branches, both correctly predict an abrupt change in the magnon dispersion around this value of $\bq$. This is due to  the Stoner spin-flip transitions having comparable energy to the magnon energy causing strong interference and intensity suppression at these values of $\bq$ (as can be seen later in Fig. \ref{f:NiChi0T}).

In contrast to Ni, the experimental dispersion for fcc Co obtained by Balashov et al. \cite{Balashov} along [100] does not show any optical branches. Both ALDA and AGGA behave the same and show good agreement with experiment, with AGGA being slightly lower in energy. Observing the full transverse response function over the whole BZ (Figs. \ref{f:NiCoFeChiT}(c) and \ref{f:NiCoFeChiT}(d)) we again see a reduction in the peak strength and suppression of the magnons by AGGA as compared to ALDA. Qualitatively, AGGA also reproduces the jump in magnon energy witnessed in experiments around $|\bq|=0.6$, although at a higher value of $|\bq|(=0.8)$.

For Fe, we see significant broadening in the AGGA (Fig. \ref{f:NiCoFeChiT}(f)) transverse response for $|\bq|>0.5$, to such an extent that it becomes impossible to assign a peak position. This may explain the experimental observation that magnons were not observed for $|\bq|>0.46$. At this value of $\bq$, we observed, with AGGA, strong suppression of the magnon due to interaction with the Stoner continuum. 

We now test the predictive power of LDA and GGA by comparing their behavior for Ni, Co, and Fe at the experimental and optimized lattice parameters (see Fig. \ref{f:NiCoFea0}). We find that the AGGA magnon spectra are more sensitive to the lattice parameters than the ALDA. In most cases, the AGGA results at the corresponding GGA parameter $(a_0)_{theo}$ are lower in energy and closer to the experimental results than at $(a_0)_{exp}$, although still overestimated. Suprisingly, for the case of Co, the GGA result is better than the LDA result.

Next we investigate Heusler (or half-Heusler) materials Co$_2$MnSi and NiMnSb, which due to their geometry of interlocking magnetic fcc lattices, can (in principle) have multiple magnon branches. In Fig. \ref{f:Heuslers} the magnon spectra of Co$_2$MnSi and NiMnSb are plotted along with the experimental and ALDA calculations. For Co$_2$MnSi (see Fig. \ref{f:Heuslers}(a)) both, an acoustic branch and an optical branch are observed. An increase in the energies of acoustic branch is noted when compared with earlier ALDA results \cite{BEBS09}, based on loss function. However, energies of the optical branch with AGGA are within the same range as reported with ALDA. One interesting aspect is that the upper branch, reported with ALDA in literature, was not seen with ALDA in this work, rather it is seen with AGGA.  
For NiMnSb (Fig. \ref{f:Heuslers}(b)), both ALDA and AGGA predict only an acoustic branch, as is also the case experimentally. This is likely due to Ni not possessing a strong local moment as most of the total moment is localized on the Mn atoms. In this case AGGA severely overestimates the magnon energies.

Finally, we offer the underlying explanation as to why AGGA tends to overestimate the magnon energies.
The role of the XC kernel is to transform the excitation structure of Im$\{\chi_0^{-+}\}$ into the true response. From the Stoner single-particle excitations, contained in Im$\{\chi_0^{-+}\}$, it must create the magnon peak.
At $\bq=0$, the gap in Im$\{\chi_0^{-+}\}$ is related to the exchange splitting between spin-up and spin-down states. This splitting dictates the position of the Stoner continuum across the BZ. In Figs. \ref{f:NiChi0T}(a) and (b), we plot Im$\{\chi_0^{-+}\}$ for LDA and GGA, where we observe that the Stoner gap has increased by approximately $60$ meV for Nickel. This increment stems from the fact that GGA increases the exchange splitting in Ni by $59.9$ meV compared to LDA, leading to the shift in Stoner continuum towards high energies. Similar behavior was also observed for other materials, e.g for Fe there is $\approx 150$ meV increase, and even a $50\%$ increase of the LDA Stoner gap in $\chi_0^{-+}$ for the half-metal NiMnSb. This increase in the Stoner excitation energies translates into an increase in the magnon energies. Given that LDA is well-known to overestimate the exchange splitting, the further enhancement on going from LDA to GGA leads to large overestimation of the magnon energies.

\section{CONCLUDING REMARKS}

We have studied the charge and spin excitation spectra using the gradient dependent AGGA XC kernel within the linear response regime of TDDFT. The calculated EELS for LiF and diamond show that the AGGA kernel performs slightly better than the ALDA kernel, although, as would be expected, neither captures excitonic effects. For magnon dispersions, AGGA generally overestimates the magnon energies. This is due to the fact that the GGA XC functional overestimates the exchange splitting. Furthermore, the intensity of the peaks is greatly suppressed in the spectra obtained by the AGGA XC kernel due to interaction of spin-waves with the Stoner continuum. This suppression is also observed in experiments, suggesting AGGA might provide better qualitative understanding than ALDA. Heusler materials consisting of multiple magnetic sublattices were also studied where it was found that AGGA is better at resolving higher-energy optical magnon branches. When experimental reference data for the system geometry and lattice parameters is unavailable, we found that using GGA consistently, i.e. for both the ground-state (i.e. lattice parameter as well as density) as well as to calculate the response function, gave better results than using the experimental parameters.  

In this work, we principally investigated the spin-spin response of collinear ferromagnetic systems, which greatly simplified the XC kernel. However, the AGGA XC kernel derived here is valid for all systems, and, in particular, has interesting terms for the the spin-charge response and for non-collinear systems. This will be explored in future work.

It is important to test the performance of adiabatic functionals in TDDFT as their behavior can be quite different from the ground-state case. Only by implementing, assessing, and understanding this behavior can we gain insight into the relevant features necessary for accurate XC kernels, which can guide us in developing new approximations in TDDFT.

\medskip

\bibliography{references} 

\onecolumngrid
\section{Appendix A}

 For the spin polarized case, the Exchange-Correlation (XC) energy functional, $E\xc$, depends on spin-up, $\rho_\alpha(\br)$, spin-down, $\rho_\beta(\br)$, densities and their gradients, $\bn \rho_\alpha(\br), \bn \rho_\beta(\br)$. The XC potential, $v\xc$, and the kernel, $f\xc$, can be obtained by the first and second order functional derivative of $E\xc$ with respect to the densities.
 
Now adding variation in the two densities and their gradients, and Taylor expanding one gets the XC energy density, $e\xc$, (up to first order only). 

\begin{equation}
\label{eq1}
\begin{split}
e\xc(\rho_\alpha(\br)+\delta \rho_\alpha(\br) , \rho_\beta(\br) &+ \delta \rho_\beta(\br), \bn^r \rho_\alpha(\br)+ \bn^r \delta \rho_\alpha(\br), \bn^r \rho_\beta(\br)+  \bn^r \delta \rho_\beta(\br)) \\ 
&= e\xc(\rho_\alpha(\br),\rho_\beta(\br),\bn^r \rho_\alpha(\br), \bn^r \rho_\beta(\br)) 
+ \dfrac{\partial e\xc}{\partial \rho_\alpha} (\br) \delta \rho_\alpha(\br) 
+  \dfrac{\partial e\xc}{\partial \rho_\beta} (\br) \delta \rho_\beta(\br) \\
&+  \dfrac{\partial e\xc}{\partial \bn \rho_\alpha} (\br) \bn^r \delta \rho_\alpha(\br)
+  \dfrac{\partial e\xc}{\partial \bn \rho_\beta} (\br) \bn^r \delta \rho_\beta(\br) 
\end{split}
\end{equation}
and 
\begin{equation}
\begin{split} \nonumber
E\xc[\rho_\alpha(\br)+\delta \rho_\alpha(\br) , \rho_\beta(\br) &+\delta \rho_\beta(\br), \bn^r \rho_\alpha(\br)+ \bn^r \delta  \rho_\alpha(\br), \bn^r \rho_\beta(\br)+  \bn^r \delta \rho_\beta(\br)] \\
&= E\xc[\rho_\alpha(\br),\rho_\beta(\br),\bn^r \rho_\alpha(\br), \bn^r \rho_\beta(\br)] \\
&+ \int d^3r \Big[ \dfrac{\partial e\xc}{\partial \rho_\alpha} (\br) \delta \rho_\alpha(\br) 
+  \dfrac{\partial e\xc}{\partial \rho_\beta} (\br) \delta \rho_\beta(\br) 
+  \dfrac{\partial e\xc}{\partial \bn \rho_\alpha} (\br)  \bn^r \delta \rho_\alpha(\br)
+  \dfrac{\partial e\xc}{\partial \bn \rho_\beta} (\br) \bn^r \delta \rho_\beta(\br) \Big] 
\end{split} 
\end{equation}
\begin{equation}
\begin{split} 
\label{dexc}
\delta E\xc =  \int d^3r \Big[ \dfrac{\partial e\xc}{\partial \rho_\alpha} (\br) \delta \rho_\alpha(\br) 
+  \dfrac{\partial e\xc}{\partial \rho_\beta} (\br) \delta \rho_\beta(\br) 
+  \dfrac{\partial e\xc}{\partial \bn \rho_\alpha} (\br)  \bn^r \delta \rho_\alpha(\br)
+  \dfrac{\partial e\xc}{\partial \bn \rho_\beta} (\br)  \bn^r \delta \rho_\beta(\br) \Big] 
\end{split}
\end{equation}

As we know

\begin{equation}
\begin{split}
\delta E\xc
&= \int d^3r v\xc^\alpha (\br) \delta \rho_\alpha (\br) 
+ \int d^3r v\xc^\beta (\br) \delta \rho_\beta (\br)
+ \frac{1}{2} \int \int d^3r d^3r' f^{\alpha\alpha}\xc(\br,\br') \delta \rho_\alpha(\br) \delta \rho_\alpha (\br') \\
&+ \frac{1}{2} \int \int d^3r d^3r' f^{\alpha\beta}\xc(\br,\br') \delta \rho_\beta(\br) \delta \rho_\alpha(\br') 
+ \frac{1}{2} \int \int d^3r d^3r' f^{\beta\alpha}\xc(\br,\br') \delta \rho_\alpha(\br) \delta \rho_\beta(\br') \\
&+ \frac{1}{2} \int \int d^3r d^3r' f^{\beta\beta}\xc(\br,\br') \delta \rho_\beta(\br) \delta \rho_\beta(\br') ,
\end{split}
\end{equation}

so we could either expand Eq. (\ref{eq1}) to 2$^{nd}$ order or use $v\xc^{\alpha,\beta}$, where $f^{\alpha\alpha}\xc$ is the change in $v^{\alpha}\xc$ when $\rho_\alpha$ changes and $f^{\alpha\beta}\xc$ is change in $v^\alpha\xc$ when $\rho_\beta$ changes, and similarly for the other spin channel. Using integration by parts in Eq.(\ref{dexc}), we obtain the XC potential for the two spin densities as:

\begin{equation}
v^{\alpha}\xc[\rho_\alpha,\rho_\beta,\bn \rho_\alpha,\bn \rho_\beta](\br)
= \dfrac{\partial e\xc}{\partial \rho_\alpha}(\br)  
- \bn_j^r \Big[ \dfrac{\partial e\xc}{\partial \bn_j \rho_\alpha}(\br)  \Big]
\end{equation}

and

\begin{equation}
v^{\beta}\xc[\rho_\alpha,\rho_\beta,\bn \rho_\alpha,\bn \rho_\beta](\br)
= \dfrac{\partial e\xc}{\partial \rho_\beta}  (\br)
- \bn_j^r \Big[ \dfrac{\partial e\xc}{\partial \bn_j \rho_\beta}(\br)  \Big]
\end{equation}

Now variation of $v^\alpha\xc$ w.r.t. $\rho$ gives the kernels $f\xc^{\alpha\alpha}(\br,\br')$ and $f\xc^{\alpha\beta}(\br,\br')$ as

\begin{equation}
\begin{split}
\delta & v^{\alpha}\xc (\br) 
= v^{\alpha}\xc[\rho_\alpha(\br)+\delta \rho_\alpha(\br) , \rho_\beta(\br) + \delta \rho_\beta(\br), \bn^r \rho_\alpha(\br)+ \bn^r \delta \rho_\alpha(\br), \bn^r \rho_\beta(\br) +  \bn^r \delta \rho_\beta(\br)]  
- v^{\alpha}\xc[\rho_\alpha,\rho_\beta,\bn \rho_\alpha,\bn \rho_\beta] \\
&= \Big\{ \frac{\partial^2 e\xc}{\partial \rho_\alpha^2} (\br) \delta \rho_\alpha (\br) 
+ \Big( \frac{\partial^2 e\xc}{\partial \bn_k \rho_\alpha \partial \rho_\alpha} (\br) \Big) \bn_k^r \delta \rho_\alpha (\br)  
- \bn^r_j \Big[ \Big( \frac{\partial^2 e\xc}{\partial \rho_\alpha \partial \bn_j \rho_\alpha} (\br) \Big) \delta \rho_\alpha (\br)  \Big] \\
&- \bn^r_j \Big[ \Big( \frac{\partial^2 e\xc}{\partial \bn_k \rho_\alpha \partial \bn_j \rho_\alpha} (\br) \Big)  \bn^r_k \delta \rho_\alpha (\br) \Big] \Big\} 
+ \Big\{ \frac{\partial^2 e\xc}{\partial \rho_\beta \partial \rho_\alpha} (\br) \delta \rho_\beta (\br)
+ \Big( \frac{\partial^2 e\xc}{\partial \bn_k \rho_\beta \partial \rho_\alpha} (\br) \Big) \bn_k \delta  \rho_\beta (\br) \\
&- \bn^r_j \Big[ \Big( \frac{\partial^2 e\xc}{\partial \rho_\beta \partial \bn_j \rho_\alpha} (\br) \Big) \delta \rho_\beta (\br) \Big] 
- \bn^r_j \Big[ \Big( \frac{\partial^2 e\xc}{\partial \bn_k \rho_\beta \partial \bn_j \rho_\alpha} (\br) \Big) \bn^r_k \delta \rho_\beta (\br) \Big] \Big\} \\
&= \int  f\xc^{\alpha\alpha}(\br,\br') \delta \rho_\alpha(\br') d^3r'
+\int  f\xc^{\alpha\beta}(\br,\br') \delta \rho_\alpha(\br') d^3r' 
\end{split}
\end{equation}

Solving each term separately by introducing an integration with a delta function, we obtain the kernels $f\xc^{\alpha\alpha}(\br,\br')$ and $f\xc^{\alpha\beta}(\br,\br')$ 

\begin{equation}
f^{\alpha\alpha}\xc(\br,\br') = \delta(\br-\br') \Big[ \frac{\partial^2 e_{xc}}{\partial \rho_\alpha \partial \rho_\alpha} (\br)
- \Big( \bn^r_k \frac{\partial^2 e_{xc}}{\partial \bn_k \rho_\alpha \partial \rho_\alpha} (\br) \Big) \Big]
- \bn^{r'}_k \Big[ [\bn^{r'}_j \delta(\br-\br')]  \Big( \frac{\partial^2 e_{xc}}{\partial \bn_k \rho_\alpha \partial \bn_j \rho_\alpha} (\br') \Big) \Big]  
\end{equation}

\begin{equation}
\begin{split}
f^{\alpha\beta}\xc(\br,\br') &= \delta(\br-\br') \Big\{ \frac{\partial^2 e\xc}{\partial \rho_\beta \partial \rho_\alpha} (\br') \Big\}
- \Big\{ [\bn^{r'}_j \delta(\br-\br')]  \Big( \frac{\partial^2 e\xc}{\partial \rho_\alpha \partial \bn_j \rho_\beta }  (\br) \Big) \Big\} \\
&+ \Big\{ [\bn^{r'}_j \delta(\br-\br')]  \Big( \frac{\partial^2 e\xc}{\partial \rho_\beta \partial \bn_j \rho_\alpha}  (\br') \Big) \Big\}
- \bn^{r'}_k \Big\{ [\bn^{r'}_j \delta(\br-\br')]  \Big( \frac{\partial^2 e\xc}{\partial \bn_k \rho_\beta \partial \bn_j \rho_\alpha} (\br') \Big) \Big\} .
\end{split} 
\end{equation}

For non-collinear systems, the kernel comprises charge-charge, $f\xc^{00}$, charge-spin, $f\xc^{0i}$, and spin-spin, $f\xc^{ij}$, terms which consist of the above $f\xc^{\alpha\alpha}$ and $f\xc^{\alpha\beta}$ terms:

\begin{equation}
\label{f00}
f\xc^{00} = \dfrac{1}{4} \Big[ f\xc^{\uparrow\uparrow} (\br,\br') 
+ f\xc^{\uparrow\downarrow} (\br,\br') 
+ f\xc^{\downarrow\uparrow} (\br,\br') 
+ f\xc^{\downarrow\downarrow} (\br,\br') \Big]
\end{equation}

\begin{equation}
\label{foi}
f\xc^{0i} = \dfrac{1}{4} \Big[ f\xc^{\uparrow\uparrow} (\br,\br') 
- f\xc^{\uparrow\downarrow} (\br,\br') 
+ f\xc^{\downarrow\uparrow} (\br,\br') 
- f\xc^{\downarrow\downarrow} (\br,\br') \Big] \hat{m}_i(\br')
\end{equation}

\begin{equation}
\label{fij}
f\xc^{ij} = \dfrac{1}{4} \Big[ f\xc^{\uparrow\uparrow} (\br,\br') 
- f\xc^{\uparrow\downarrow} (\br,\br') 
- f\xc^{\downarrow\uparrow} (\br,\br') 
+ f\xc^{\downarrow\downarrow} (\br,\br') 
- \dfrac{|\textbf{B}\xc|}{|\textbf{m}|} \Big] \hat{m}_i(\br) \hat{m}_j(\br') 
+ \dfrac{|\textbf{B}\xc|}{|\textbf{m}|} I_3
\end{equation}

where $I_3$ is the $3\times3$ Identity matrix, $\hat{m}$ is the unit magnetization vector, $|\textbf{B}\xc|$ is the magnitude of magnetic field, and $|\textbf{m}|$ is the magnitude of the magnetization.

\end{document}